# Acoustic quasi-lossless transmission in arbitrary pathway of network

Hongqing Dai, Tingting Liu, Baizhan Xia [a)], Dejie Yu [b)]

(State Key Laboratory of Advanced Design and Manufacturing for Vehicle Body, Hunan University, Changsha, Hunan, People's Republic of China, 410082)

Acoustic metamaterials have exhibited extraordinary possibilities to manipulate the propagation of the sound wave. Up to now, it is still a challenge to control the propagation of the sound wave in an arbitrary pathway of a network. Here, we design a symmetry breaking cross-shape metamaterial comprised of Helmholtz resonant cells and a square column. The square column is eccentrically arranged. The sound wave can quasi-lossless transmit through channels along the eccentric direction with compressed spaces, which breaks through the general transmission phenomenon. This exotic propagation characteristic is verified by the band structure and the mode of the metamaterial. Two acoustic networks, including a 2x2 network and an 8x8 network, show that the sound wave can quasi-lossless propagate along various arbitrary pathways, such as the Great Wall shape, the stair step shape and the serpentine shape, by reconfiguring eccentric directions. This ability opens up a new venue to route the sound wave and exhibits promising applications, from the acoustic communication to the energy transmission.

---

[a)] E-mail: xiabz2013@hnu.edu.cn (Baizhan Xia)

[b)] djyu@hnu.edu.cn (Dejie Yu)

The acoustic metamaterial is a periodically distributed artificial structure with the ability to manipulate the propagation of the sound wave, and has presented a promising application such as acoustic cloaking[1-3], supertunneling[4,5], superlensing[6-8] or logic gate[9,10]. The important characteristic of the acoustic metamaterial is its forbidden band gap. Up to now, there are two main mechanisms responsible for the generation of the forbidden band gap. One is the Bragg scattering in the composite material with periodic variations of the material density and the elastic modulus. The other one is the local resonance[11], which greatly promotes the development of the locally resonant acoustic metamaterial. The common locally resonant acoustic metamaterials are the Helmholtz resonator[12-15], the tensioned membrane[16-19] and the mass-and-spring system[20,21]. It is worth noted that an entirely novel class of electromagnetic metamaterial with the topological edge state has been recently developed[22,23]. Inspired by the electronic edge states occurring in the topological insulators, the concept of the topological acoustic has been proposed[24] and the phenomenon of disorder-free one-way sound propagation has been demonstrated[25]. What's more, a phononic metamaterial with a topologically protected elastic wave that was robust against scattering from discrete defects and disorders was constructed[26]. These developments inspired us that we could try to control the elastic wave in an arbitrary pathway by utilizing acoustic metamaterials.

Lossless transmission has important applications in signal transmissions[27]. Due to the existence of the energy dissipation, a lossless transmission seems to be impossible. But the extremely small transmission loss, namely the quasi-lossless transmission, can be achievable[28-30]. In order to realize the quasi-lossless transmission, an electromagnetic metamaterial with super-tunneling, namely the epsilon-near-zero metamaterial was presented[31]. It was demonstrated that the epsilon-near-zero metamaterial can be used to squeeze the electromagnetic energy through a narrow subwavelength waveguide. Subsequently, experiments were carried out to demonstrate that electromagnetic metamaterials with an electric permittivity near zero could present pass gaps with high efficient transmissions[32,33]. Analogous to electromagnetic metamaterials, acoustic metamaterials with supertunneling have also been developed[4,5,34]. The energy transmission of acoustic wave through a narrow channel filled with metamaterials can be up to 87.5%, while it will be reduced to 3% for the case without metamaterials[34]. It indicates that the acoustic metamaterial has a great promising application value in the high efficient transmission and provides a wonderful opportunity for the quasi-lossless transmission.

Symmetry breaking is a significant approach that has been utilized to design the extraordinary metamaterial[35,36]. By breaking the symmetry of the coupled split-ring resonator system, one could excite a dark resonant mode that was not easily accessible in a symmetric split-ring resonator structure[37]. Similarly, by breaking the four-fold rotational symmetry of the tetragonal $URu_2Si_2$, an in-plane anisotropy of the magnetic susceptibility emerged. These two significant discoveries exhibit that the symmetry breaking is conducive to investigate the "hidden" properties of metamaterials. Spontaneous symmetry breaking in torsional chiral magneto elastic structures can lead to a giant nonlinear polarization change, energy localization and mode splitting, and provides a new possibility to create an artificial phase transition in metamaterials[38]. Recently, a semi-analytical model breaking the symmetry of the spatial arrangement of single resonant cells was presented[8]. It was demonstrated that a single negative metamaterial could be turned into a double negative one by judiciously breaking its symmetry. More recently, some researchers discovered that the negative permeability band could be enlarged by breaking the structural symmetry of a conventional cut-wire-pair metamaterial[39]. Inspired by the above explorations of symmetry breaking, we intend to design a symmetry breaking metamaterial for the quasi-lossless transmission of the sound wave in an arbitrary pathway of network.

In this paper, we design a cross-shape metamaterial comprises of Helmholtz resonant cells and a square column. The symmetry of the cross-shape metamaterial is broken due to the eccentric arrangement of the square column. We calculate the transmission properties of the symmetry breaking cross-shape metamaterial by employing the finite element method. The results show that the sound wave can transmit through channels along the eccentric direction with compressed spaces, which breaks through the general transmission phenomenon. We also investigate two acoustic networks (namely a 2x2 network and an 8x8 network) consisting of cross-shape metamaterials. The results show that the eccentric cross-shape metamaterial can produce an excellent propagation of the sound wave in an arbitrary pathway of a network.

# Results and discussions

**Symmetry breaking acoustic metamaterial.** Fig. 1a shows a cross-shape metamaterial with a centric square steel column. The side length of the square cavity is $6d$. The length and width of the rectangular channel are respectively $4d$ and $2d$. A square column is inserted in the center of the cross-shape metamaterial, and the length of the square is $l_0$. In order to break the symmetry of the cross-shape metamaterial, we replace the centric square column with an eccentric one, as shown in Fig. 1b. The length of the square is $l_e$. The red arrow represents the eccentric direction. Due to the eccentric arrangement of the square column, the top space and the right space of the square cavity are compressed. The square cavity and four rectangular channels are filled with uniformly distributed unit cells. The unit cell is consisted of 8 triangular Helmholtz cavities, as shown in Fig. 1c. The periodic constant of the unit cell is $d$. The side length of the unit cell is $a$. Width and length of the short neck of the Helmholtz resonator are respectively $b$ and $c$. Thickness of the wall between two adjacent Helmholtz resonators is $t$. The unit cell is made of metal that can be assumed as rigid.

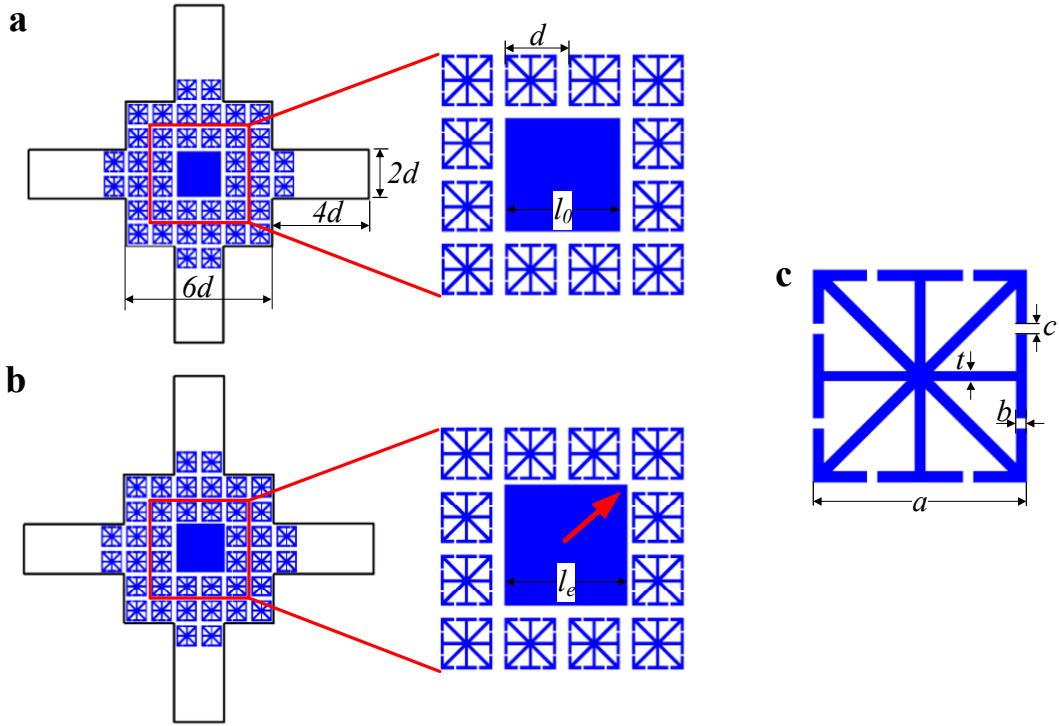

**Figure 1 | Schematic diagrams of cross-shape metamaterials.** (**a**) The cross-shape metamaterial with a centric square steel column. $l_0$=45mm, $d$=25mm. (**b**) The cross-shape metamaterial with an eccentric square column. The red arrows represent the eccentric direction. $l_e$=48mm. (**c**) The unit cell. $a$=20mm, $b$=1mm, $c$=1mm and $t$=1mm.

**Transmission characteristic of the symmetry breaking cross-shape metamaterial.** As shown in Fig. 2a and 2b, because of the symmetry of geometric structures, the transmission coefficients in the left and right channels are equivalent. Fig. 2b shows that for the symmetrical cross-shape metamaterial, the transmission coefficients of the left, the bottom and the right channels in the frequency range [3980Hz, 4030Hz] are much less than 1. Namely, the sound pressure almost cannot transmit from the top channel to the other three ones. For the symmetry breaking cross-shape tube without unit cells, it can be found from Fig. 2c that the transmission coefficients of the right and left channels are not equivalent because of the eccentric arrangement of the square column. The transmission coefficient of the sound wave from the top channel to the right one is less than the transmission coefficient to the left one, as the right space is narrower than the left one. For the symmetry breaking cross-shape metamaterial, Fig. 2d shows that the transmission coefficients of the left and bottom channels in the frequency range [3980Hz, 4030Hz] are much less than 1. Excitingly, the transmission coefficient of the right channel sharply increases. At frequencies 3995Hz and 4026Hz, the transmission coefficient exhibits two peaks with values 97.7% and 98.8%. Namely, the sound wave can almost perfectly transmit from the top channel to the right one at 3995Hz and 4026Hz. Similar to Fig. 2c, Fig. 2e shows that without unit cells, the sound wave can transmits from the left channel to the other three ones with less transmission coefficients. Furthermore, the transmission coefficient of the sound wave from the left channel to the top one is less than the transmission coefficient to the bottom one, as the top space is narrower than the bottom one. Fig. 2f shows that the transmission coefficients of the top, bottom and right channels in the frequency range [3980Hz, 4030Hz] are much less than 1. It indicates that the sound wave almost cannot transmit from the left channel to the other three ones.

From the discussion presented above, we can obtain that the symmetry breaking cross-shape metamaterial can efficiently transmit sound energy between the top and right channels. Namely, with the support of the symmetry breaking cross-shape metamaterial, the sound wave can well transmit through channels with compressed spaces along the eccentric direction. Conversely, without metamaterials, the transmission coefficient of the sound wave transmitting through channels with larger spaces is larger than that with compressed spaces. Thus, the symmetry breaking cross-shape metamaterial breaks with the general transmission characteristics of the sound wave.

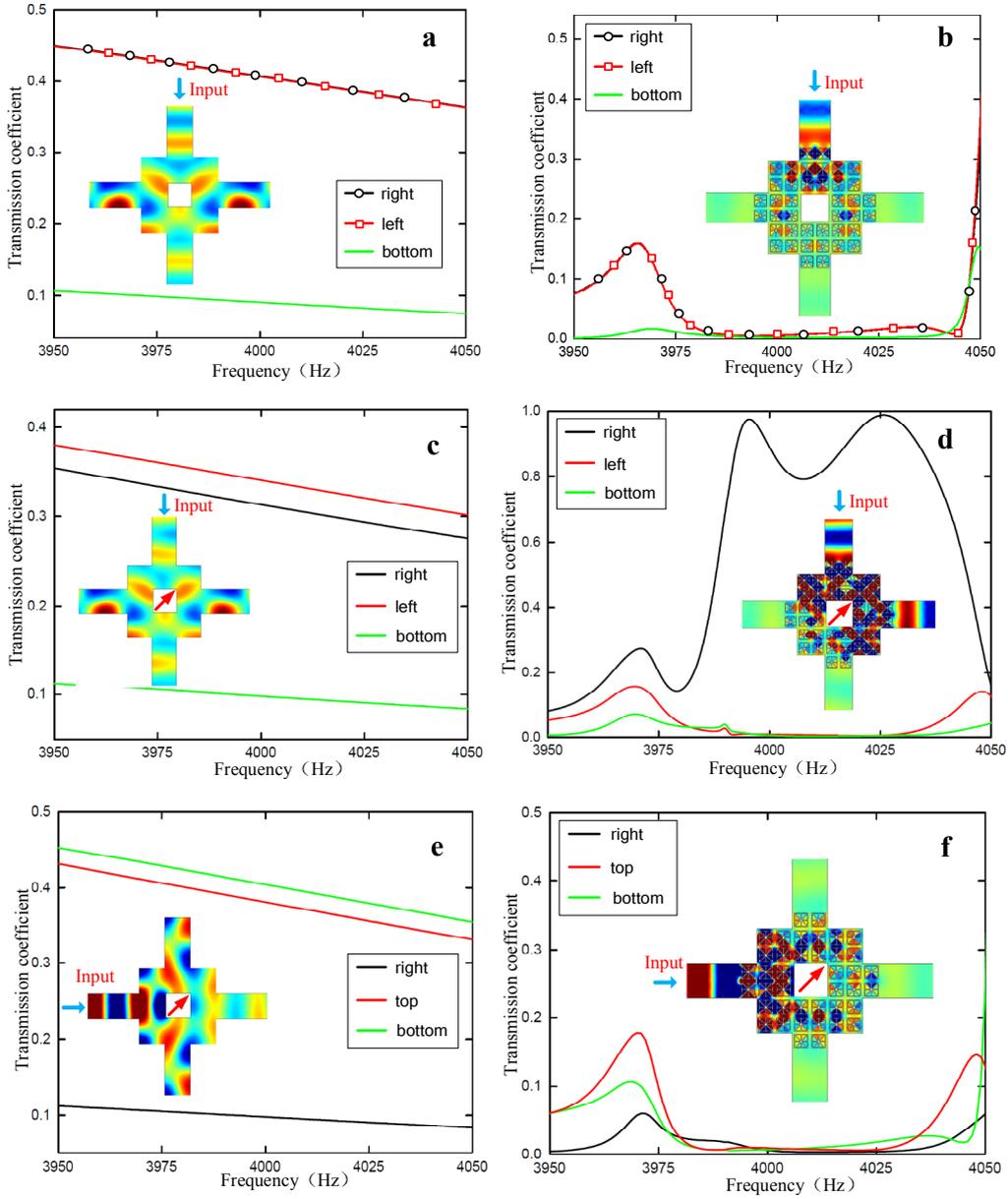

**Figure 2 | Transmission coefficients and sound pressure field distributions at 3995Hz. The red arrow represents the eccentric direction.** (**a**) and (**b**) Transmission coefficients of the symmetrical cross-shape tube without resonant unit cell and the symmetrical cross-shape metamaterial. The sound wave is incident from the top channel. (**c**) and (**d**) Transmission coefficients of the symmetry breaking cross-shape tube without resonant unit cell and the symmetry breaking cross-shape metamaterial. The sound wave is incident from the top channel. (**e**) and (**f**) The sound wave is incident from the left channel. The sound pressure field distributions of six cases at 3995Hz are plotted in (a)-(f)

**Modal analysis of the symmetry breaking cross-shape metamaterial.** In order to explore the mechanism of the extraordinary transmission characteristics, we calculate the mode of the symmetry breaking cross-shape metamaterial. As shown in Fig. 3, the symmetry breaking cross-shape metamaterial has two modes at 3989Hz and 4021Hz. In both modes, we can note that the top channel and the right channel are interconnected. It indicates that the sound wave can freely transmit between the top channel and the right channel. However, except of this pair of channels (top-right), the propagation of the sound wave between the other pairs of channels (such as top-down, top-left, left-top, et al.) is truncated. This phenomenon is tallied with the results presented in Fig. 2.

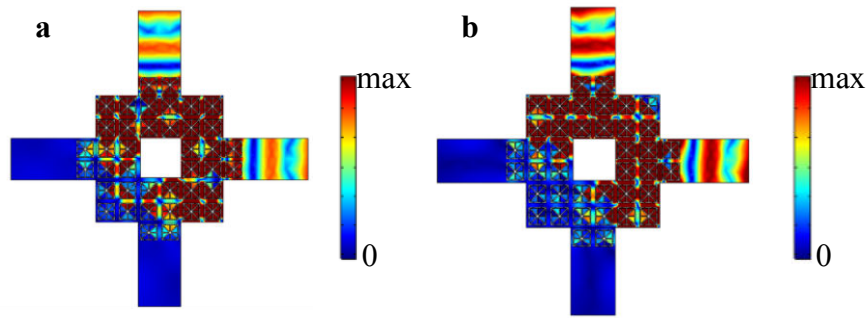

Figure 3 | Modes of the symmetry breaking cross-shape metamaterial at 3989Hz (a) and at 4021Hz (b).

**The band analysis of the symmetry breaking cross-shape metamaterial.** The band structures of the centric cross-shape metamaterial and the symmetry breaking cross-shape metamaterial are presented in Fig. 4. As shown in Fig. 4a, there are two complete band gaps [3966Hz, 4005Hz] and [4014Hz, 4038Hz] for the centric cross-shape metamaterial. As shown in Fig. 4b, there are five narrow band gaps [3968Hz, 3979Hz], [3982Hz, 3995Hz], [3999Hz, 4017Hz], [4021Hz, 4028Hz], and [4033Hz, 4039Hz] for the symmetry breaking cross-shape metamaterial. By comparing Fig. 4b with Fig. 4a, we can find that the band gaps of the centric cross-shape metamaterial are broken, and two new pass gaps represented by the red lines are generated in the symmetry breaking cross-shape metamaterial. Namely, in the first band gap [3966Hz, 4005Hz] (in Fig. 4a), a new pass gap [3995Hz, 3999Hz] (in Fig. 4b) is yielded. In the second band gap [4014Hz, 4038Hz] (in Fig. 4a), a new pass gap [4017Hz, 4021Hz] (in Fig. 4b) is yielded. In both pass gaps, the sound wave can transmit with a high efficiency. It should be noted that both pass gaps are in the vicinity of peaks of the transmission coefficient.

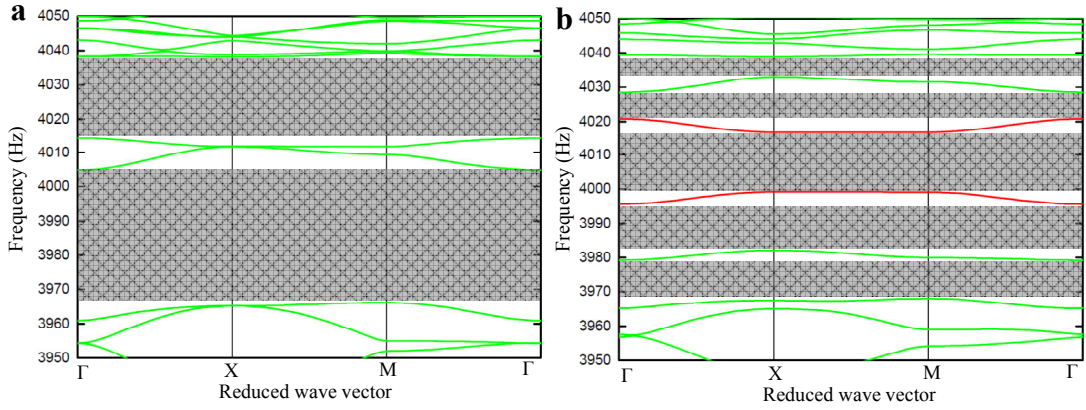

**Figure 4 | The band structures of the centric cross-shape metamaterial (a) and the symmetry breaking cross-shape metamaterial (b).**

**Transmission property of a 2x2 network.** In order to further explore the transmission property of the symmetry breaking cross-shape metamaterial, we have performed the numerical simulation of a 2x2 network. The channels of the network are numbered from 1 to 8 in an anticlockwise direction, as shown in Fig. 5. By modulating eccentric directions, the sound wave can transmit through various channels. When the eccentric direction of the metamaterial 1 is the upper-right corner, its top and right channels are interconnected. When the eccentric direction of the metamaterial 2 is the left-bottom corner, its left and bottom channels are interconnected. When the eccentric direction of the metamaterial 3 is top-left corner, its top and left channels are interconnected. When the eccentric direction of the metamaterial 4 is the right-bottom corner, its right and bottom channels are interconnected. In this case, the sound wave can transmit from the channel 1 to the channel 4 by passing through all metamaterials, as shown in Fig. 5a. When the eccentric direction of the metamaterial 3 is the top-right corner, its top and right channels are interconnected. In this case, the sound wave cannot transmit to the metamaterial 4. Instead, the sound wave will transmit to the channel 6, as shown in Fig. 5b. When the eccentric direction of the metamaterial 2 is the left-top corner, its left and top channels are interconnected. In this case, the sound wave cannot transmit to the metamaterial 3. Instead, the sound wave will transmit to the channel 8, as shown in Fig. 5c. When the eccentric direction of the metamaterial 1 is the upper-left corner, its top and left channels are interconnected. In this case, the sound wave cannot transmit to the metamaterial 2. Instead, the sound wave will transmit to the channel 2, as shown in Fig. 5d. Results in Fig. 5 indicate that with a deliberate arrangement of eccentric square column, the sound wave can transmit from an odd channel to any even channel, and vice versa.

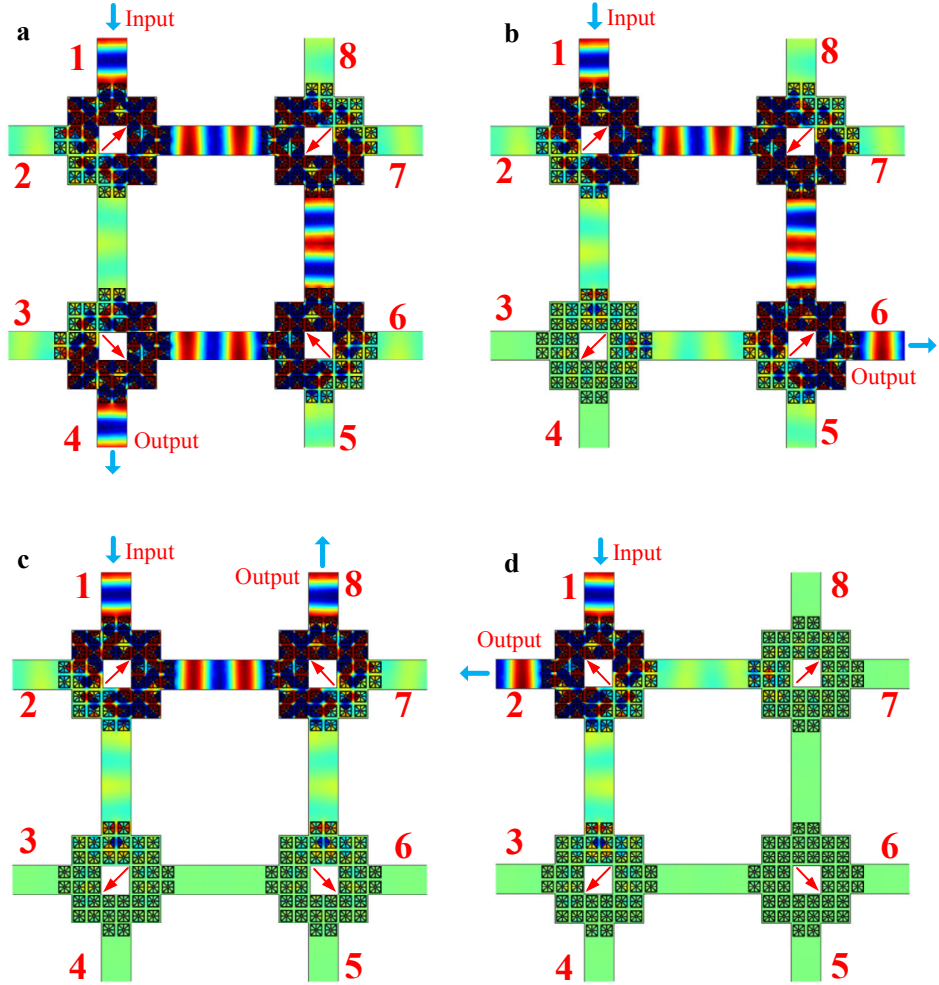

**Figure 5 | Sound pressure field distributions of a 2x2 network at 3995Hz. The red arrows represent the eccentric directions. The sound wave is incident from the channel 1.** (**a**) The export is the channel 4. (**b**) The export is the channel 6. (**c**) The export is the channel 8. (**d**) The export is the channel 2.

**Transmission property of an 8x8 network.** An 8x8 network consisted of sixty-four components is investigated. The quasi-lossless transmission of the symmetry breaking cross-shape metamaterial allows an ideal reflection-less routing along arbitrarily defined pathways, by simply modulating the square column. To confirm the possibility of such arbitrary routing, we have generated three typical pathways shown in Fig. 6. Fig. 6a shows that the sound wave can propagate in the acoustic network along the Great Wall shape pathway. Fig. 6b presents that the sound wave can propagate in the acoustic network along the diagonal pathway, like stair steps. Fig. 6c presents that the sound wave can propagate in the acoustic network along the serpentine pathway. Indeed, the pathway of the sound wave can be further reconfigured by modulating eccentric directions.

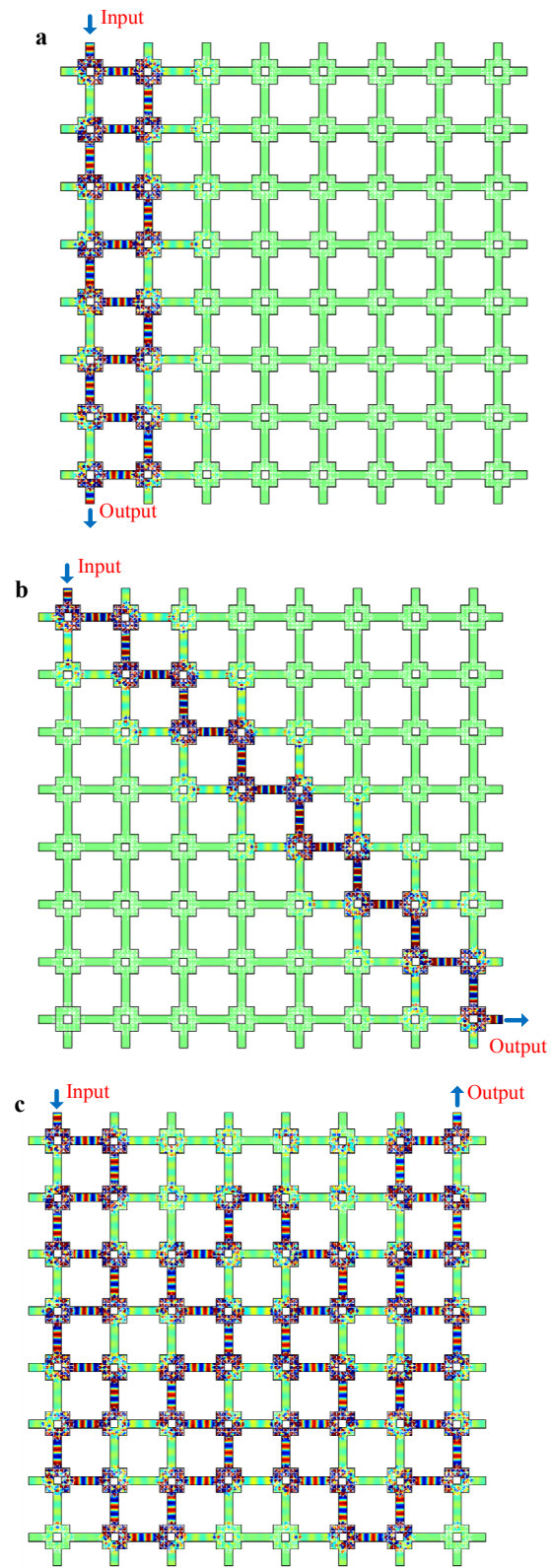

**Figure 6 | Sound pressure field distributions of an 8x8 network at 3995Hz.** (**a**) Pathway with the Great Wall shape. (**b**) Pathway with the stair steps shape. (**c**) Pathway with the very serpentine shape.

**Conclusion**

To conclude, we design a symmetry breaking cross-shape metamaterial to achieve quasi-lossless transmission in an arbitrary pathway of a network. It demonstrates that the sound wave can propagate through channels along the eccentric direction with compressed spaces, which breaks with the general transmission phenomenon. Two simulations, including a 2x2 network and an 8x8 network, show that the symmetry breaking cross-shape metamaterial has the ability to produce a quasi-lossless propagation of the sound wave along arbitrarily defined pathway by reconfiguring eccentric directions. Based on these outstanding properties, we can envision unprecedented potentialities to route the sound wave, achieving excellent propagation characteristics. We believe that quasi-lossless transmission concepts can enormously expand the engineering toolkit of modern acoustic devices, and open up a new versatile venue to control the propagation of the sound wave.

**Method**

**Finite elements simulations.** The simulation was performed with the commercial finite element analysis solver Comsol Multiphysics. The condition of narrow region acoustics in acoustic pressure module was selected to investigate the propagation characteristics. The incident sound pressure is a plane wave. The boundary conditions of the channels were set to the plane wave radiation boundary. The mode and band structures were also obtained by solving the eigenvalue problem derived from the finite element analysis solver Comsol Multiphysics. The boundary conditions of the cross-shape metamaterials were set to the Floquet periodicity conditions.

**References**

1. Popa, B. I., Zigoneanu, L. & Cummer, S. A. Experimental Acoustic Ground Cloak in Air. *Phys. Rev. Lett.* **106**, 253901 (2011).
2. Zhu, X., Liang, B., Kan, W., Zou, X. & Cheng, J. Acoustic cloaking by a superlens with single-negative materials. *Phys. Rev. Lett.* **106**, 014301 (2011).

3. Zigoneanu, L., Popa, B. I. & Cummer, S. A. Three-dimensional broadband omnidirectional acoustic ground cloak. *Nat. Mater.* **13**, 352-355 (2014).

4. Liang, Z. & Li, J. Extreme acoustic metamaterial by coiling up space. *Phys. Rev. Lett.* **108**, 114301 (2012).

5. Graciá-Salgado, R., García-Chocano, V. M., Torrent, D. & Sánchez-Dehesa, J. Negative mass density and ρ-near-zero quasi-two-dimensional metamaterials: design and applications. *Phys. Rev. B* **88**, 224305 (2013).

6. Shen, C., Xu, J., Fang, N. X. & Jing, Y. Anisotropic Complementary Acoustic Metamaterial for Canceling out Aberrating Layers. *Phys. Rev. X* **4**, 041033 (2014).

7. Wang, W. Q., Xie, Y. B., Konneker, A., Popa, B. I. & Cummer, S. A. Design and demonstration of broadband thin planar diffractive acoustic lenses. *Appl. Phys. Lett.* **105**, 101904 (2014).

8. Kaina, N., Lemoult, F., Fink, M. & Lerosey, G. Negative refractive index and acoustic superlens from multiple scattering in single negative metamaterials. *Nature* **525**, 77 (2015).

9. Zhang, T., Cheng, Y., Guo, J.-z., Xu, J.-y. & Liu, X.-j. Acoustic logic gates and Boolean operation based on self-collimating acoustic beams. *Appl. Phys. Lett.* **106**, 113503 (2015).

10. Zhang, T., Cheng, Y., Yuan, B.-G., Guo, J.-Z. & Liu, X.-J. Compact transformable acoustic logic gates for broadband complex Boolean operations based on density-near-zero metamaterials. *Applied Physics Letters* **108**, 183508 (2016).

11. Liu, Z. *et al.* Locally resonant sonic materials. *Science* **289**, 1734-1736 (2000).

12. Fang, N. *et al.* Ultrasonic metamaterials with negative modulus. *Nat. Mater.* **5**, 452-456 (2006).

13. Hu, X., Ho, K.-M., Chan, C. & Zi, J. Homogenization of acoustic metamaterials of Helmholtz resonators in fluid. *Phys. Rev. B* **77**, 172301 (2008).

14. García-Chocano, V., Graciá-Salgado, R., Torrent, D., Cervera, F. & Sánchez-Dehesa, J. Quasi-two-dimensional acoustic metamaterial with negative bulk modulus. *Phys. Rev. B* **85**, 184102 (2012).

15. Park, C. M. & Lee, S. H. Propagation of acoustic waves in a metamaterial with a refractive index of near zero. *Appl. Phys. Lett.* **102**, 241906 (2013).

16. Yang, Z., Mei, J., Yang, M., Chan, N. & Sheng, P. Membrane-type acoustic metamaterial with negative dynamic mass. *Phys. Rev. Lett.* **101**, 204301 (2008).

17. Park, C. M. *et al.* Amplification of acoustic evanescent waves using metamaterial slabs. *Phys. Rev. Lett.* **107**, 194301 (2011).

18. Zhu, R., Liu, X. N., Huang, G. L., Huang, H. H. & Sun, C. T. Microstructural design and experimental validation of elastic metamaterial plates with anisotropic mass density. *Phys. Rev. B* **86**, 144307 (2012).

19. Xiao, S., Ma, G., Li, Y., Yang, Z. & Sheng, P. Active control of membrane-type acoustic metamaterial by electric field. *Appl. Phys. Lett.* **106**, 091904 (2015).

20. Huang, H. H. & Sun, C. T. Theoretical investigation of the behavior of an acoustic metamaterial with extreme Young's modulus. *J. Mech. Phys. Solids* **59**, 2070-2081 (2011).

21. Huang, H. H. & Sun, C. T. Anomalous wave propagation in a one-dimensional acoustic metamaterial having simultaneously negative mass density and Young's modulus. *J. Acoust. Soc. Am.* **132**, 2887-2895 (2012).

22. Qi, X.-L. & Zhang, S.-C. Topological insulators and superconductors. *Rev. Mod. Phys.* **83**, 1057-1110 (2011).

23. Hasan, M. Z. & Kane, C. L. Colloquium: Topological insulators. *Rev. Mod. Phys.* **82**, 3045-3067 (2010).

24. Yang, Z. *et al.* Topological acoustics. *Phys. Rev. Lett.* **114**, 114301 (2015).

25. Wang, P., Lu, L. & Bertoldi, K. Topological Phononic Crystals with One-Way Elastic Edge Waves. *Phys. Rev. Lett.* **115**, 104302 (2015).

26. Mousavi, S. H., Khanikaev, A. B. & Wang, Z. Topologically protected elastic waves in phononic metamaterials. *Nat. Commun.* **6**, 8682 (2015).

27. Aniacastañón, J. Quasi-lossless transmission using second-order Raman amplification and fibre Bragg gratings. *Opt. Express* **12**, 4372-4377 (2004).

28. Ellingham, T. J., Ania-Castanon, J. D., Ibbotson, R. & Chen, X. Quasi-lossless optical links for broad-band transmission and data processing. *IEEE Photonics Technol. Lett.* **18**, 268-270 (2006).

29. Karalekas, V. *et al.* Performance optimization of ultra-long Raman laser cavities for quasi-lossless transmission links. *Opt. Commun.* **277**, 214-218 (2007).

30. Jia, X. H. *et al.* Detailed theoretical investigation on improved quasi-lossless transmission using third-order Raman amplification based on ultralong fiber lasers. *J. Opt. Soc. Am. B* **29**, 847-854 (2012).

31. Silveirinha, M. & Engheta, N. Tunneling of electromagnetic energy through subwavelength channels and bends using epsilon-near-zero materials. *Phys. Rev. Lett.* **97**, 157403 (2006).

32. Edwards, B., Alu, A., Young, M. E., Silveirinha, M. & Engheta, N. Experimental verification of epsilon-near-zero metamaterial coupling and energy squeezing using a microwave waveguide. *Phys. Rev. Lett.* **100**, 033903 (2008).

33. Liu, R. *et al.* Experimental demonstration of electromagnetic tunneling through an epsilon-near-zero metamaterial at microwave frequencies. *Phys. Rev. Lett.* **100**, 023903 (2008).

34. Cheng, Y. *et al.* Ultra-sparse metasurface for high reflection of low-frequency sound based on artificial Mie resonances. *Nat. Mater.* **14**, 1013 (2015).

35. Kante, B. *et al.* Symmetry breaking and optical negative index of closed nanorings. *Nat. Commun.* **3**, 1180 (2012).

36. Biancoli, A., Fancher, C. M., Jones, J. L. & Damjanovic, D. Breaking of macroscopic centric symmetry in paraelectric phases of ferroelectric materials and implications for flexoelectricity. *Nat. Mater.* **14**, 224-229 (2015).

37. Aydin, K., Pryce, I. M. & Atwater, H. A. Symmetry breaking and strong coupling in planar optical metamaterials. *Opt. Express* **18**, 13407-13417 (2010).

38. Liu, M., Powell, D. A., Shadrivov, I. V., Lapine, M. & Kivshar, Y. S. Spontaneous chiral symmetry breaking in metamaterials. *Nat. Commun.* **5**, 4441 (2014).

39. Trang, P. T. *et al.* Symmetry-Breaking Metamaterials Enabling Broadband Negative Permeability. *J. Electron. Mater.* **45**, 2547-2552 (2016).

**Acknowledgments**

The paper is supported by National Natural Science Foundation of China (No.11402083, 11572121).

**Author contributions**

All authors contributed extensively to the work presented in this paper. B.Xia and D.Yu conceived the idea, H.Dai and T.Liu performed the numerical simulations, H.Dai and B.Xia prepared the manuscript, with comment and correction from D.Yu.